\documentclass[twocolumn,amsmath,amssymb,prx]{revtex4-1}

\usepackage{graphicx}
\usepackage{dcolumn}
\usepackage{bm}
\usepackage{color}

\begin{document}

\title{Quantifying Structural Dynamic Heterogeneity in a Dense Two-dimensional Equilibrium Liquid}

\author{Tamoghna Das}
\email{dtamoghna@me.com}
\affiliation{
Center for Nanoscale Science and Technology, National Institute of Standards and Technology, Gaithersburg, MD 20899, USA
}
\affiliation{
Maryland Nanocenter, University of Maryland, College Park, MD 20742, USA.
}
\author{Jack F. Douglas}
\email{jack.douglas@nist.gov}
\affiliation{
Materials Science and Engineering Division, National Institute of Standards and Technology, Gaithersburg, Maryland 20899, USA.
}


\begin{abstract}
We investigate local structural fluctuations in a model equilibrium fluid with the aim of better understanding the structural basis of locally heterogeneous dynamics identified in recent simulations and experimental studies of glass-forming liquids and other strongly interacting particle systems, such as lipid membranes, dusty plasmas, interfacial dynamics of crystals, the internal dynamics of proteins, etc. In particular, we utilize molecular dynamics simulation methods to study single component Lennard-Jones condensed material at constant temperature in two dimensions (2D) over a range of densities covering both the liquid and crystalline phase regimes. We identify three distinct structural classes of particles by examining the immediate neighborhood of individual particles relying on a solid-angle based tessellation technique. The area distribution of the neighborhoods reveals cages having hexagonal, pentagonal and square symmetries. Pentagonal cells appear to be the predominant motif in the liquid phase, while the solid phase is dominated by hexagonal cells, as in the case of a perfect crystal. An examination of the spatial organization of particles belonging to each structural class further indicates that finite-size clusters of the hexagonal and pentagonal particle populations arise within both liquids and solids and the size of these clusters grow in a complementary way as a function of density. Both particle populations form percolation clusters in liquid-crystal coexistence regime. Interestingly, the populations of particles with different local structures, defined by the arrangement of neighboring particles, are found to maintain different diffusivities, as computed from the velocity autocorrelation function for each type of particles for all densities studied. Our analysis provides a new conceptual framework for understanding the structural origin of dynamical heterogeneity in soft materials.
\end{abstract}

\maketitle

\section{Introduction}
The structure of a liquid is generally perceived as a constantly evolving, irregular spatial arrangement of particles or molecules. \cite{LBook1,LBook2,LBook3} The combined characteristics of short-range local correlations associated with local molecular packing and the lack of long range order make the structural characterization of liquids a difficult task. \cite{Bernal1, Bernal2} Over the past century, several theoretical mean field approaches such as the integral equation theory, \cite{inteqn1, inteqn2, inteqn3} density functional theory, \cite{dft} free volume theory, \cite{fv1, fv2, fv3, fv4} mode coupling theory, \cite{mct1, mct2, mct3}  and the Adam-Gibbs model of glass-formation \cite{AG} have been introduced to understand the thermodynamic and dynamic properties of cooled dense liquids, but it is fair to conclude that that a unified and predictive model of the thermodynamics and dynamics of liquids remains elusive. 

Starting from the seminal work of Rahman, \cite{rahman} it has become increasingly appreciated that local fluid structural fluctuations which are averaged out in the calculation of the pair correlation function, must be important for understanding the dynamics of strongly interacting fluids. Shape fluctuations of the Voronoi cells around individual fluid atoms, investigated by Rahman, were later found to be strongly correlated with the direction of atomic displacement, \cite{fail2, vt5} a finding that served to define collective coordinates defining the atomic displacements of individual atoms. This picture of atomic motion connects up with Barker tunnel model \cite{tunnel1} in which atoms in liquids are conceived to move in `tunnels' -- defined by their interaction with the surrounding molecules, a model allowing for analytic estimation of both thermodynamic and dynamic properties of fluids. \cite{tunnel2, tunnel3, tunnel4} The work of Rahman and Barker anticipates  the stringlike motion in cooled liquids emphasized in recent modelling of glass-forming liquids. \cite{mobimob, dw3} Although Rahman's approach never developed into a full-fledged theory of the dynamics and thermodynamics of fluids, we think that the underlying philosophy of his work remains promising, and accordingly, we employ a tessellation method to define local structural fluctuations in a model 2D equilibrium fluid, which allows for easy visualization of local structural heterogeneities and the large scale organization of clusters of these structural heterogeneities in space. Our goal is to identify structural counterparts to particles having three distinct dynamical stated observed previously in simulation: `mobile', `immobile' and `normal' particles \cite{mobimob}  and the large scale organization of particles representing each class of particles.

Our search for a structural basis for this dynamic heterogeneity (DH) has numerous antecedents, many of which have ended in failure. Specifically, there is no demonstrated general link at present between any kind of local structural heterogeneity and dynamic heterogeneity in liquids. For example, the volume of Voronoi cells, a measure of the local fluid density, is a poor predictor of local mobility so that the widely invoked and intuitively attractive `free volume' model relating local fluid density to local fluid mobility is simply not correct. \cite{fail1} If a structural indicator of local mobility exists in fluids, then this quantity is a more subtle property than local density. More recent works aimed at characterizing structural heterogeneity in fluids have emphasized possible root causes of such heterogeneity: icosahedral clustering of atoms, \cite{icos1, icos2, icos3} the formation of medium-range ordered regions \cite{mro1, mro2} and abstract `defect' structures in dense fluids. Indeed, locally preferred packing arrangements in fluids has been shown to exist and to be correlated with particles having low mobility in some glass-forming systems. \cite{slow1, slow2, slow3, slow4} Unfortunately, these structural motifs do not arise universally in all glass-formers \cite{nonu1, nonu2} and the relation between these structural heterogeneity and resulting dynamic properties such as the molecular diffusion coefficient and structural relaxation time remains unclear. `Defects' in liquids might conceivably be defined in terms of some type of loosely-packed regions, but the disordered nature of the liquid state makes it difficult to precisely define any kind of defect structure. 
        
Now if we follow in the footsteps of Rahman and define local neighborhoods and structural fluctuations in terms of a tessellation, we need a method that is computationally efficient, experimentally realizable and capable of defining structural signatures of significance in relation to previously observed changes in local atomic mobility. The Voronoi tessellation, or its dual, the Delaunay triangulation method, can unequivocally identify nearest-neighbor particles for uniformly distributed points representing particle locations in both two and three dimensions. \cite{vtbook1} For ideal crystals, the Voronoi cell patterns reduce to the Wigner-Seitz tessellation, \cite{SolidBook2} representing the `reciprocal lattice' of the crystal. While no such parallel physical interpretation of the Voronoi tessellation can be drawn for liquids, this construction leads to an appealing description of liquids in terms of a dynamical foam-like `structure' defined by the geometry of the tessellation cells rather than the positions of the dual particles. The Voronoi tessellation, though the most popular choice, is not an obvious one for spatially inhomogeneous systems such as liquids and fluid mixtures. \cite{vtbad} Pan et al. \cite{cna} have discussed the shortcomings of the Vornoi tessellation for modeling the structural heterogeneity of fluids and have introduced a modified Voronoi method that better quantifies changes in the effective coordination number in fluids through an elaborate `common neighbor analysis'. A downside of this interesting method is that it is rather computational expensive and somewhat difficult to implement in experimental studies. 

The present work is based on another tessellation method, the {\em relative angular distance} (RAD) tessellation method, \cite{rad} which is relatively simple to implement both experimentally and computationally and which, similarly to the modified Voronoi method adopted in Ref.\cite{cna}, emphasizes variations in the local effective coordination number, a property recently correlated in recent machine learning studies with variations in local mobility in liquids. \cite{machinelearning} We show below that our tessellation method is well-adapted to identifying subclasses of particles defining distinct structural environments and dynamics, although the construction does not emphasize dislocation and disclination structures observed conspicuously in a Voronoi tessellation of melting two-dimensional crystals. It is not clear whether this is a drawback or advantage of the new tessellation method. In our first study of this tessellation methodology, we consider an equilibrium two-dimensional fluid since it is obviously easier to visualize the tessellation geometry for this geometry.  There should be no difficulty formally in developing the construction in higher dimensions and treating particles of different sizes and shapes. We mention that our analysis of structural correlations in model two-dimensional fluids appears to be a complement to the aperture cross-correlation functional analysis of Sheu and Rice, \cite{q2D1, q2D2} but the study of these different approaches to quantifying local structural fluctuations requires further investigation.

The organization of the rest of the paper is as follows. Sec. 2 introduces the model system, describes the simulation method and presents a primary characterization of the system. In Sec. 3, the neighbor identfication algorithm is elaborated and one quantfication of the fluctuating neighborhood is described in detail. Further analysis based on this tessellation is presented in Sec. 4 in three parts: (i) different structural populations are identified by our tessellation method, (ii) structural correlations of the particles in each population and (iii) the average dynamical properties for each structural class are quantified. Our main findings are summarized and their implications are discussed in Sec. 5.

\section{Model system and simulation details}
A system of $N$ mono-disperse particles of size $\sigma$ and unit mass within a square box with periodic boundary condition is considered as a model two-dimensional system. The particles interact pair-wise via a Lennard-Jones interaction, $V(r) = 4\epsilon\left[(\sigma/r)^{12}-(\sigma/r)^6\right]$ where $r=|{\bf r}_i-{\bf r}_j|$ is the distance between any arbitrary pair $\{i,j\}$ of particles within the system. $\epsilon$ and $\sigma$ set the energy and length scales of the system, respectively, and both of them are set to unity. $V(r)$ asymptotically approaches to zero value at $r_c=2.5\sigma$, which is set as a global cut-off of interaction. Each particle experience a conservative force ${\bf F}_c$ from all particles within $r_c$ computed as ${\bf F}_c=V^\prime(r)\hat{\bf r}$ where {\em prime} denotes the first spatial derivative of $V(r)$ and $\hat{\bf r}$ is the pair-wise unit vector. Other than ${\bf F}_c$, particles are subjected to two different pairwise forces, a disipative force ${\bf F}_d=-\gamma_d w_d(r) (\hat{\bf r}\cdot{\bf v})\hat{\bf r}$ and a random force ${\bf F}_r=\gamma_r w_r(r)\theta_{ij}\hat{\bf r}$. $\theta_{ij}$ represents a Gaussian random fluctuation with zero mean and unit variance, chosen independently for each pair $i,j$ of interacting particles at each time step. The dissipative part ${\bf F}_d$ is dependent the vectorial velocity difference between the particles $i$ and $j$, ${\bf v}={\bf v}_i-{\bf v}_j$ for all instants. All forces act in a way such that the total linear and angular momentum of the system is always conserved. It has been shown rigorously \cite{fdt} that the weight factors ($w$'s) and amplitudes ($\gamma$'s) of the ${\bf F}_d$ and ${\bf F}_r$ should hold the following relative dependence obeying the fluctuation-dissipation relationship: $w_d(r)=w_r^2(r)$ and $\gamma_r^2=2\gamma_d k_BT$, where $T$ is the temperature, measured in units of $\epsilon$. $k_B$ is the Boltzmann constant, conveniently set to unity. The weight factor, $w_r(r)=1-(r/r_c)$, varies between zero and unity depending on the pair-wise distance $r$. {\em Dissipative particle dynamics} (DPD) \cite{dpd} thus simulates the statistical mechanical properties of a many-body interacting particle system in a canonical constant number-area-temperature (NAT) ensemble. Since this method was originally developed to simulate  mesoscopic systems, such as homopolymer melts, employing this technique broadens the scope of applicability of our simulation. However, the usage of any other thermostats, \cite{mdbook} local (Langevin thermostat) or global (N{\'o}se-Hoover thermostat) suitable for particulate systems such as neutral colloids or nanoparticles, should not affect the estimate of equilibrium thermodynamic properties.

\begin{figure}[h!]
\begin{center}
\includegraphics[width=0.95\linewidth]{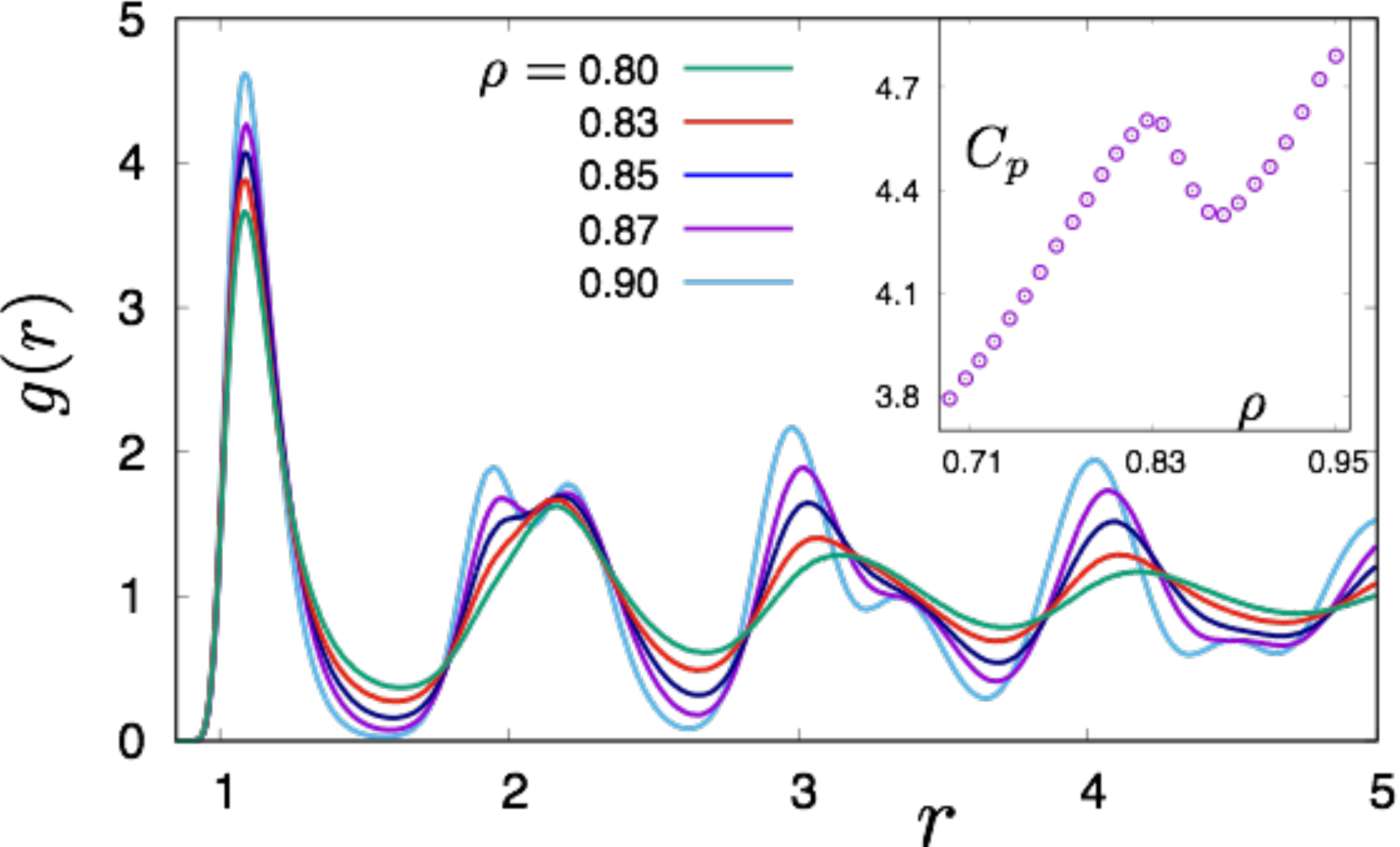}
\end{center}
\caption{
Radial distribution function, $g(r)$ is plotted as a function of distance $r$ from any arbitrary particle, for a set of densities $\rho$. $g(r)$ indicates the angularly-averaged local density around a particle as a function of distance $r$ from its centre. For $\rho=0.80$ ({\em green}) an oscillatory behavior is observed, typical to random arrangement of particles in a liquid. A shoulder on the second peak is indicative of particle crowding at $\rho=0.83$ ({\em red}). With increasing fluid density, this shoulder becomes prominent ($\rho=0.85$, {\em dark blue}) and this feature grows to the point where there is a splitting of second peak ($\rho=0.87$, {\em purple}). This peak splitting phenomenon is considered to be a signature of regions having high local ordering coexisting with relatively disordered regions. We also show $g(r)$ for $\rho=0.90$ ({\em light blue}), corresponding to the crystalline state. {\em (Inset)} Inflection of the constant pressure specific heat $C_p$ as a function of $\rho$ allows for a precise estimate of the coexistence regime. In particular, systems with $\rho<0.83$ and $\rho>0.87$ should then be considered as equilibrium liquids and solids, respectively, while intermediate densities, $0.83\le\rho\le0.87$, show the coexistence region of both phases.
}
\label{gr}
\end{figure}

Using LAMMPS, \cite{lammps} a large scale molecular dynamics simulator, we have prepared a system of $N=60,000$ particles at $k_BT=0.7\epsilon$ for a set of densities. The bulk density $\rho$ of the system is varied over a range $\rho=0.7$ to $1.0$ in increments of $0.01$ by adjusting the size of the box suitably at the beginning of the simulation. The equation of motion of each particle is solved using velocity Verlet algorithm with a time step $\Delta t=10^{-3}\tau$ where $\tau=\sqrt{m\sigma^2/\epsilon}$ is the time unit. Each system is allowed to evolve for $10^3\tau$ with a DPD thermostat, described before, for equilibration. Temperature fluctuations were measured after a period of the order of $10^{-3}$. These values remained the same for a time over two decades longer, confirming the equilibration of our model system. We then store the last $100$ configurations, separated from each other by $\tau$, for our analysis. Across the range of chosen densities, the model system is known to exhibit a liquid to solid transition through a coexistence phase. This is confirmed by computing the constant pressure specific heat $C_p$ for each $\rho$ from the fluctuation of enthalpy, $H$ as $C_p=\langle H^2\rangle-\langle H\rangle^2$. $C_p$ plotted as a function of $\rho$ (Fig.\ref{gr} {\em Inset}) shows two clear inflections at $\rho=0.83$ and $0.87$ denoting the boundaries of the coexistence phase, $0.83\le \rho\le 0.87$. Systems with $\rho<0.83$ and $\rho>0.87$ should then be considered as an equilibrium liquid and crystalline phase, respectively.

Structural support for this observation comes from the pair correlation function $g(r)$ as a function of $\rho$, which plotted in Fig.\ref{gr}. Also known as radial distribution function, $g(r)$ is one of the primary methods widely used in structural characterization of disordered media. $g(r)=\frac{1}{\rho}\langle\sum\delta(r-r_i)\rangle$, represents the average probability of finding a particle at a distance $r$ from $i$-th particle chose arbitrarily. For $\rho=0.80$, $g(r)$ shows an oscillatory behavior at small $r$, comparable to a few particle diameter, pointing to {\em short-range ordering}, typical to liquids. With increasing $r$, the oscillations die out exponentially and $g(r)$ tends asymptotically to unity. As the system enters the liquid-solid coexistence regime ($\rho=0.83$), a shoulder appears in the second peak of $g(r)$ resulting from particle crowding. With increasing $\rho$, the shoulder becomes prominent from more and more particle crowding. For $\rho=0.87$, we observe a splitting of the second peak which is indicative of particles packing in relatively ordered and disordered configurations. \cite{spltpk} The position of peaks observed for lager $\rho$ is fixed from the lattice symmetry of the crystal phase of the system. However, $g(r)$, being an angular- and ensemble-averaged quantity, does not provide a microscopic information about the structure of the material underlying this feature. Typically, the position of the first minimum of $g(r)$ is used as a cut-off distance to find the first nearest neighbors of individual particles. For systems with near crystalline order, $\rho\geq0.87$, this cut-off is found at the fixed value $r=1.5\sigma$ which yields mostly six nearest neighbors, as expected in 2D crystals. However, the cut-off shifts to larger values of $r$ as $\rho$ decreases. As a result, the distribution of nearest neighbor numbers gets broadened ranging from the lowest of $3$ nearest neighbors to maximum of upto $9$ nearest neighbors. Such $\rho$ dependence thus introduces arbitrariness in fixing the cut-off which, in turn, may lead to an ambiguous estimation of neighborhood. In the next section, we introduce a purely geometric method for nearest-neighbor identification, followed by the structural analysis based on this construction.

\section{Characterization of local neighborhood}
The relative angular distance (RAD) algorithm, \cite{rad} adapted in this work, utilizes both positional and angular information of particles in a locally adaptive way to identify the nearest-neighbors. Within a given configuration of polydisperse particles, particle $j$ is considered as a nearest neighbor to particle $i$ if $j$ and every particle closer to $i$ are {\em unblocked}. This is ensured by the following simple geometric inequality: $\Omega_{ij}>\Omega_{ik}\cos\phi_{jik}$, where $\phi_{jik}$ is the three-body angle formed at $i$ by its two prospective neighbors $j$ and $k$. The solid angle $\Omega_{ij}$, subtended by $j$ at the center of $i$ can be expressed as $\Omega_{ij}=4\pi\sin^2(\alpha_{ij}/2)$. Using a small angle approximation $\sin(\alpha_{ij}/2)\sim\sigma_{ij}/(4r_{ij})$, the inequality can be  expressed as: $(\sigma_j/r_{ij})^2>(\sigma_k/r_{ik})^2\cos\phi_{jik}$, where $r$'s are the center to center distance from $i$-th particle to particle $j$ and $k$ with size $\sigma_j$ and $\sigma_k$, respectively. For monodisperse systems, such as ours, the relation simplifies to $r^2_{ik}>r^2_{ij}\cos\phi_{jik}$. The algorithm can be implemented efficiently and non-recursively by checking the criterion for a set of nearest-neighbors sorted in ascending order of distance from the central particle.

\begin{figure}[h!]
\begin{center}
\includegraphics[width=0.95\linewidth]{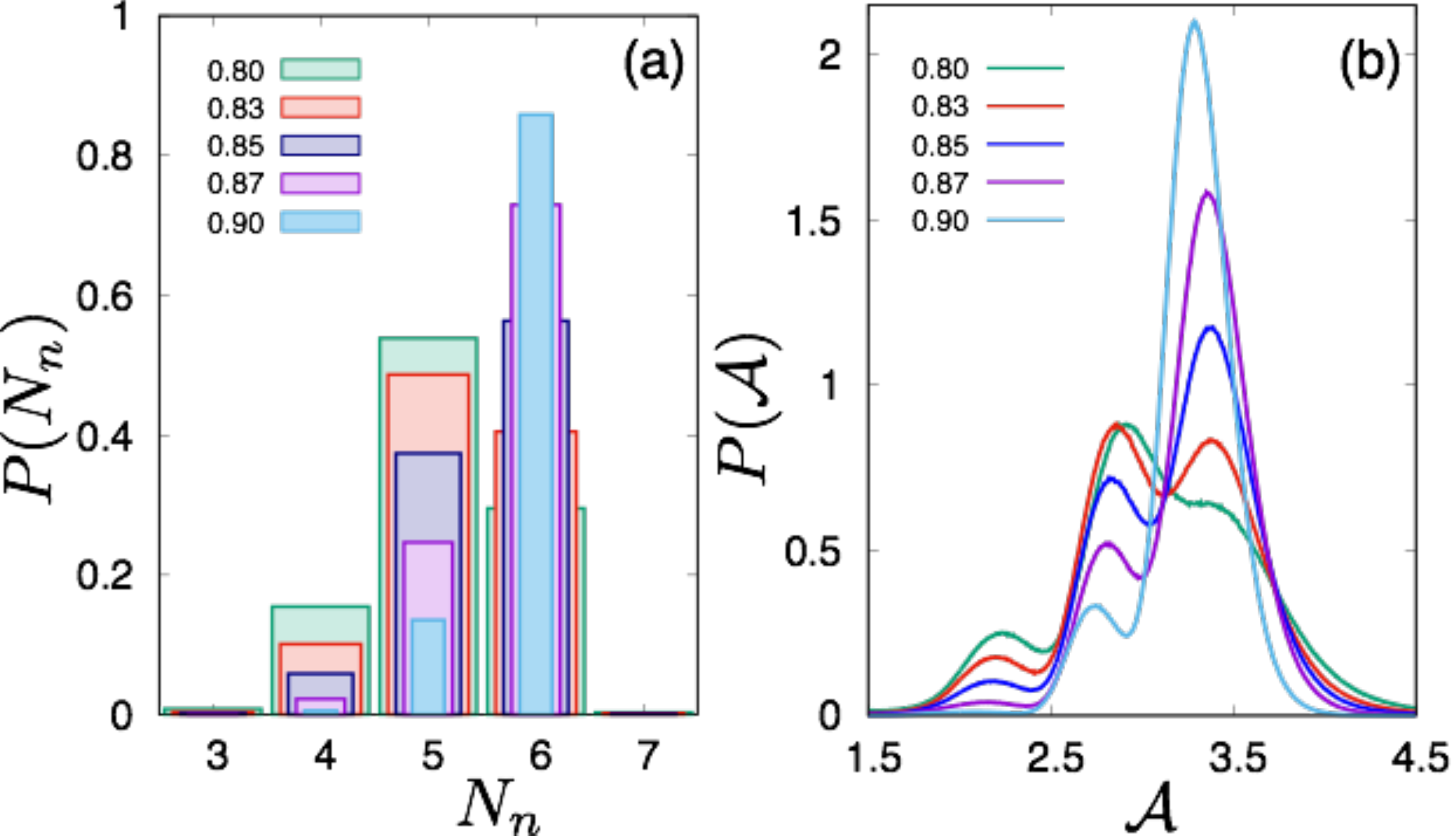}
\end{center}
\caption{
(a) Probability distribution of the number of nearest-neighbors $N_n$ of a particle for different $\rho$ is plotted. Crystalline phase at $\rho=0.90$ ({\em light blue}) is supported by the prominent peak of $P(N_n)$ at $N_n=6$. This peak decreases as the hexagonal order is destroyed with decreasing density. Particles with $N_n=5$ becomes dominant in the liquid phase, $\rho=0.80$ ({\em green}). $4$-coordinated particles also appear within liquids. (b) Normalized distribution of the area $\mathcal{A}$ enclosed by the nearest-neighbors around a particle is presented which shows a trimodal feature for all $\rho$. Particles with small area neighborhood appear to be negligible compared to particles with neighborhood of intermediate and large areas within the crystalline phase ($\rho=0.90$). In fact, the dominance of large neighborhood particles continues throughout the coexistence regime ($\rho>0.83$). Intermediate area neighborhood becomes dominant in the liquid phase ($\rho<0.83$). Keeping the $\rho$ dependence of dominant coordination numbers in mind, we safely map the small, intermediate and large area neighborhood particles to square, pentagonal and hexagonal shape structural classes.
}
\label{lklnbrhd}
\end{figure}
We start the structural analysis of our system simply by counting the number of nearest-neighbors $N_n$ identified through RAD algorithm. The distribution $P(N_n)$ for high $\rho(=0.90)$ shows that most particles are $6$-coordinated, as found in the crystalline phase where triangular packing is exhibited. As $\rho$ decreases, particles with $N_n<6$ start to appear more and more in the system. The $5$-coordinated particles become dominant over $6$-coordinated ones at $\rho=0.83$ which marks the beginning of liquid phase. While a perfect five-fold symmetric structure cannot be most the probable local neighborhood shape, the existence of such structures has been conjectured and experimental evidences are indicated in previous works. \cite{fivefold, fivefold0, fivefold1, fivefold2, fivefold3, fivefold4, fivefold5} Our finding of $5$-coordinated particles as the dominant motif of liquid structure is definitely in line with those observations. Next, we turn our attention to the area $\mathcal{A}$ around each particle enclosed by the polygons with the nearest neighbors at their vertices. The distribution $P(\mathcal{A})$ shows a consistent trimodal feature for all $\rho$ studied; see [Fig.\ref{lklnbrhd}(b)]. A weightage shift from {\em medium} area to {\em large} area is observed at $\rho>0.83$, i.e., when the system is in the regime of liquid solid coexistence. With increasing $\rho$, the {\em large} area peak dominates over the other two.

RAD based tessellation technique is advantageous over the traditional fixed cut-off method as being purely geometric, it does not suffer from the arbitrariness inherent to the latter. While this new technique offers much easier and substantially faster identification of nearest-neighbors than that is achievable by Voronoi method. The RAD tessellation method also reveals certain statistical properties of the neighborhood which are inaccessible by Voronoi method. The area (volume in three dimensions) distribution $P(\mathcal{A})$ of Voronoi neighbors are known to follow a phenomenological universal behavior \cite{vt2,vt4,vt5,fail1} of suitably scaled Gamma distribution but offers no specific physical insight about the local configuration and its change. On the other hand, $P(\mathcal{A})$ yield by RAD algorithm is found to be reliably sensitive to the liquid-solid phase transformation. Most importantly, the above mentioned trimodal feature of $P(\mathcal{A})$ promises a clean characterization of the structural heterogeneity of liquids in terms of distinct and well-defined local configurational motifs. In fact, in the next section, we present the results of such characterization in terms of local effective coordination number which enables us to study the collective organization and average dynamics of such structural motifs. Another distinction of the RAD from the Voronoi tessellation method is that dislocation defects involving pairs of 5-7 coordinated particles are not emphasized. These disclination defects play a central role in theories of melting in 2D \cite{2Dmelt0, 2Dmelt1, 2Dmelt2, 2Dmelt3, 2Dmelt4, 2Dmelt5, 2Dmelt6} and are readily observed computationally and experimentally as the melting transition is approached. These defects are not so evident in the liquid regime, however, motivating the exploration of alternate tessellation schemes that can inform on both structural and dynamic heterogeneity in non-crystalline materials. Future work should investigate how the current tessellation method performs in the regime where materials are highly crystalline. 

\section{Results}
\subsection{Identification of {\em three} geometric populations}
\begin{figure}[h!]
\begin{center}
\includegraphics[width=0.95\linewidth]{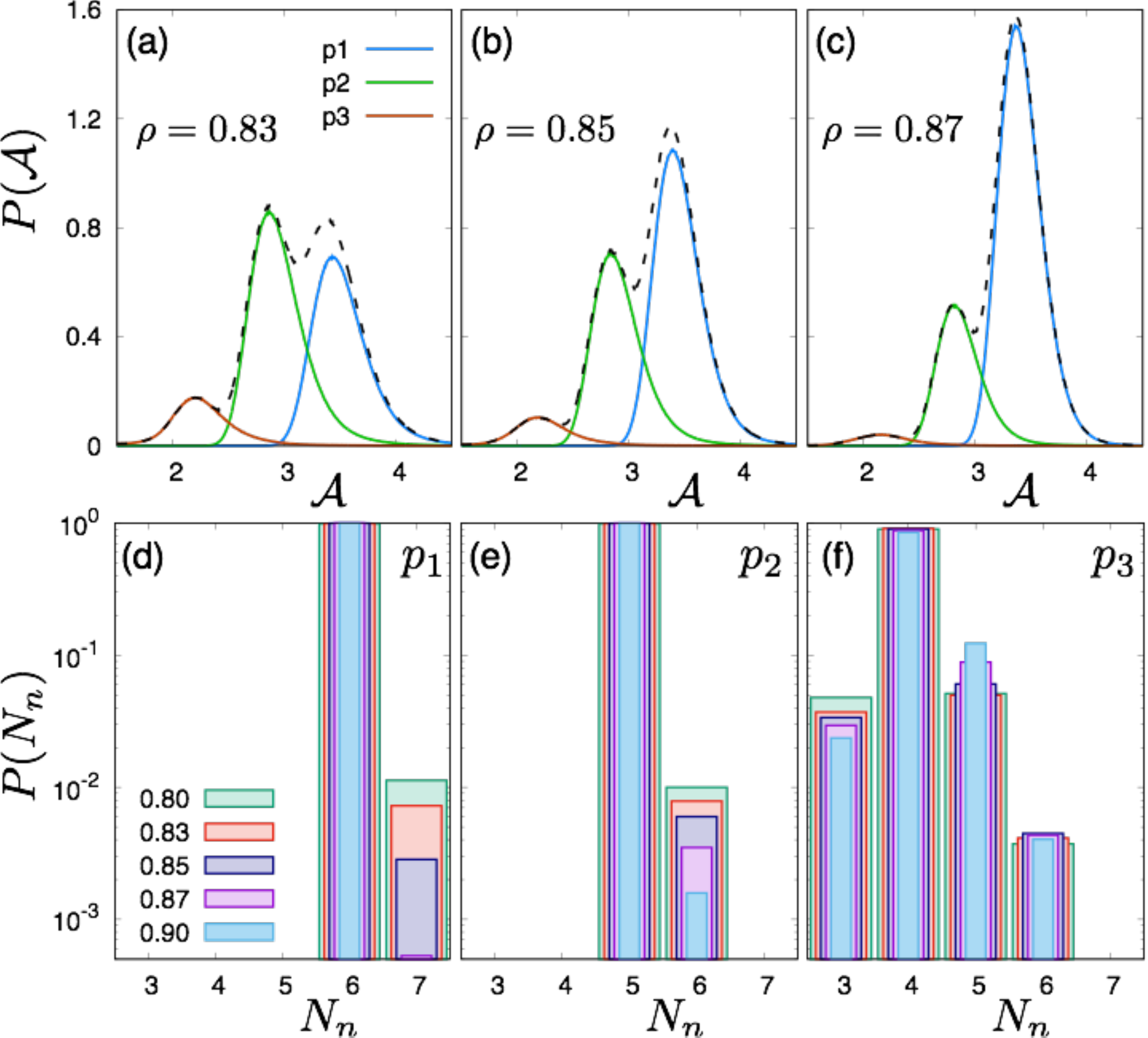}
\end{center}
\caption{
Trimodal feature of $P(\mathcal{A})$ can be reproduced by the weighted sum of three Gaussians as shown for $\rho=$ (a) $0.83$, (b) $0.85$ and (c) $0.87$. This enables us to identify three distinct statistically possible populations based on the area of individual particle neighborhood: large $p_1$ ({\em blue}), intermediate $p_2$ ({\em green}) and small $p_3$ ({\em red}). Full curves are shown by {\em dashed} lines in each of them. Sorting out the constituent particles for different populations, $P(N_n)$ is now plotted for each of the populations (d) $p_1$, (e) $p_2$ and (f) $p_3$, separately. The dominant peaks at $6, 5$ and $4$, respectively for $p_1, p_2$ and $p_3$ support our mapping between coordination number and neighborhood area. However, note that $P(N_n)$ plotted in the {\em log}-scale shows that the dominant peak is at least order of magnitude stronger than the others.
}
\label{pop}
\end{figure}
Guided initially by pure intuition, we first attempt to fit the trimodal $P(\mathcal{A})$ with three Gaussian distributions with distinct means and variances. Sum of the weight of each Gaussian is always fixed at unity. The goodness of fit for each $\rho$, as for example is shown in Fig.\ref{pop}(a), (b) and (c), enables us to identify {\em three} distinct geometric populations based on their mean $\mathcal{A}$: large area $p_1$, medium area $p_2$ and small area $p_3$. Next, we seek for a $N_n$-$\mathcal{A}$ mapping by recomputing $P(N_n)$, now, for particles of a specific population. The results of this exercise, shown in Fig.\ref{pop}(d), (e) and (f), point to the following mapping of different population: $p_1:N_n=6$, $p_2:N_n=5$ and $p_3:N_n=4$ based on the dominant $N_n$ found for each population. Since a specific $\mathcal{A}$ can be achieved by numerous particle arrangements, the mapping between number of particles and the area enclosed by them is not, in general, obvious. The statistical mapping revealed by our analysis thus provides a quantification of the degree of local structural ordering or `structural heterogeneity' which is valid for a range of $\rho$ covering liquid-solid transition of our model system. As a corollary, it suggests that a {\em coordination number description}, popularly used for molecular liquids \cite{coordL} and granular fluids, \cite{coordG} can only provide an incomplete overall description of the system. A more detailed microscopic description can be achieved through quantification of local geometry i.e., neighborhood area used in this study. Results of other geometric quantification such as the shape of particle neighborhoods, are discussed elsewhere. \cite{Lettppr}

\subsection{Collective nature of local geometric fluctuations}
\begin{figure}[h!]
\begin{center}
\includegraphics[width=0.95\linewidth]{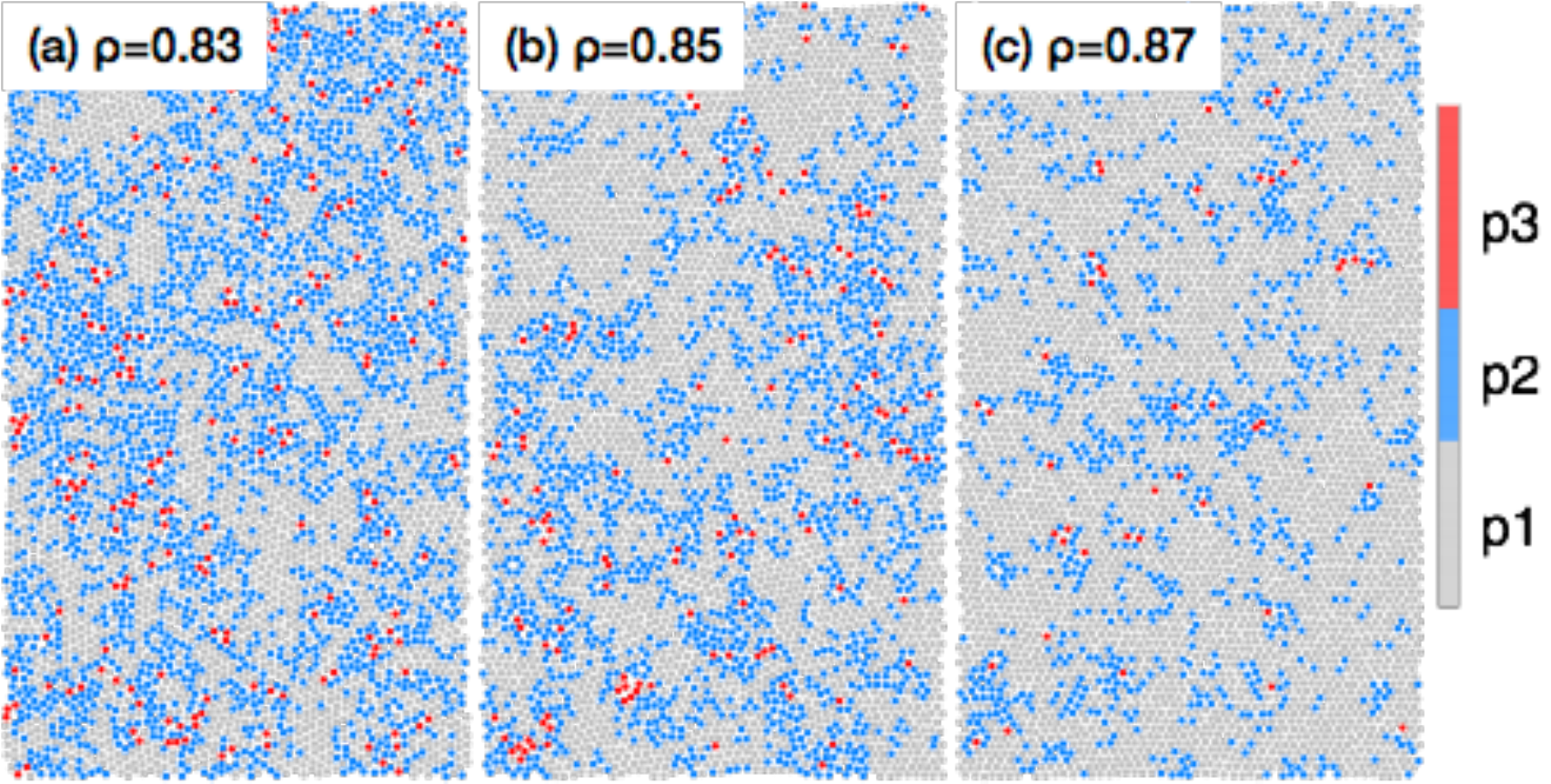}
\end{center}
\caption{
Spatial organization of particles belonging to different populations is presented from $\rho=$ (a) $0.83$, (b) $0.85$ and (c) $0.87$ with $p_1$ ({\em gray}), $p_2$ ({\em blue}) and $p_3$ ({\em red}), for each $\rho$. Particles with five-fold neighborhood, $p_2$, dominant at low-density (liquid phase) decrease with increasing $\rho$ associated to increase in $p_1$ (six-fold neighborhood). Rare abundance of $p_3$, population of particles with mostly four-fold neighborhood, increase with decreasing $\rho$. Only a small portion of the full simulation box is shown for visual clarity.
}
\label{map}
\end{figure}

Direct visualization of different population (Fig.\ref{map}) supports our previous observations: $p_2$ particles ({\em blue}) with mostly $5$ neighbors are the majority in systems with low $\rho$. $6$-coordinated particles of $p_1$ population ({\em gray}), minority in the liquid phase, gradually grows with increasing density and become dominant in the crystalline phase. Particles belonging to $p_3$ population ({\em red}) always occur in small numbers compared to other two populations which become even rare in the solid phase. Another important aspect of their spatial organization is also evident from direct visualization: particles of similar neighborhood type seem to form clusters and we next characterize this clustering phenomena.
\begin{figure}[h!]
\begin{center}
\includegraphics[width=0.75\linewidth]{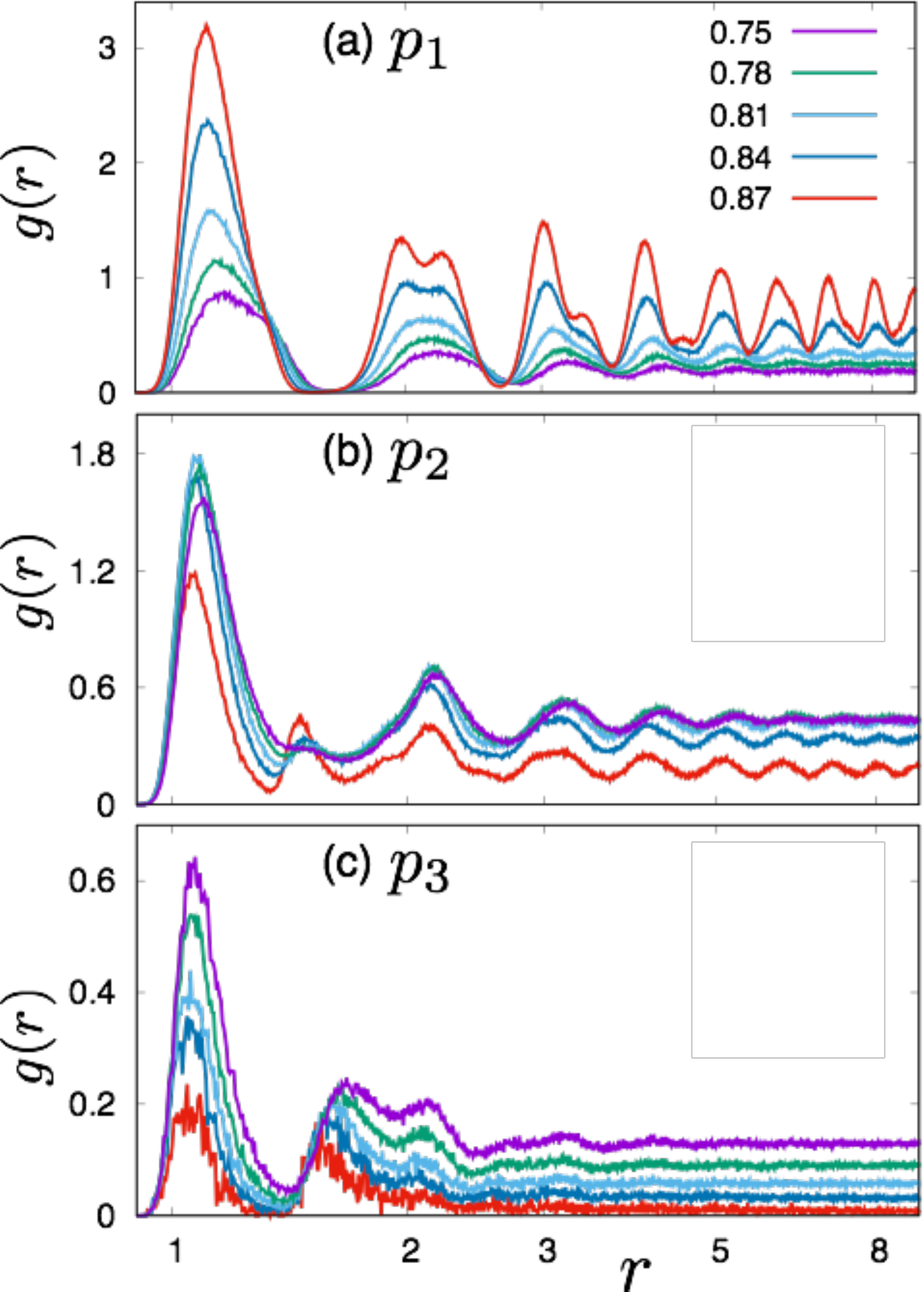}
\end{center}
\caption{
Radial distribution $g(r)$ of particles in each population is presented separately for (a) $p_1$, (b) $p_2$ and (c) $p_3$ for increasing $\rho$. (a) $g(r)$for $p_1$ shows the prominent split in second peak for high $\rho$, characteristic of a crystalline phase. At low $\rho$, it shows liquid-like periodic oscillation while maintaining its six-fold but possibly distorted neighborhood. This indicates overall structural disorder, evident from the shoulder formation of the first peak at low $\rho$. (b) Expected liquid-like features are observed for $g(r)$ of $p_2$ except appearance of a small secondary peak, close to the first peak, with increasing $\rho$. This feature, indicative of a fractal arrangement of particles, becomes abruptly prominent at $\rho\ge0.87$. (c) A dense {\em gas}-like feature is observed in $g(r)$ of $p_2$ which is expected. Appearance of more peaks at large $r$ with increasing $\rho$ asserts local organization of such particles within the system.
}
\label{grlkl}
\end{figure}
\begin{figure*}[t!]
\begin{center}
\includegraphics[width=0.85\linewidth]{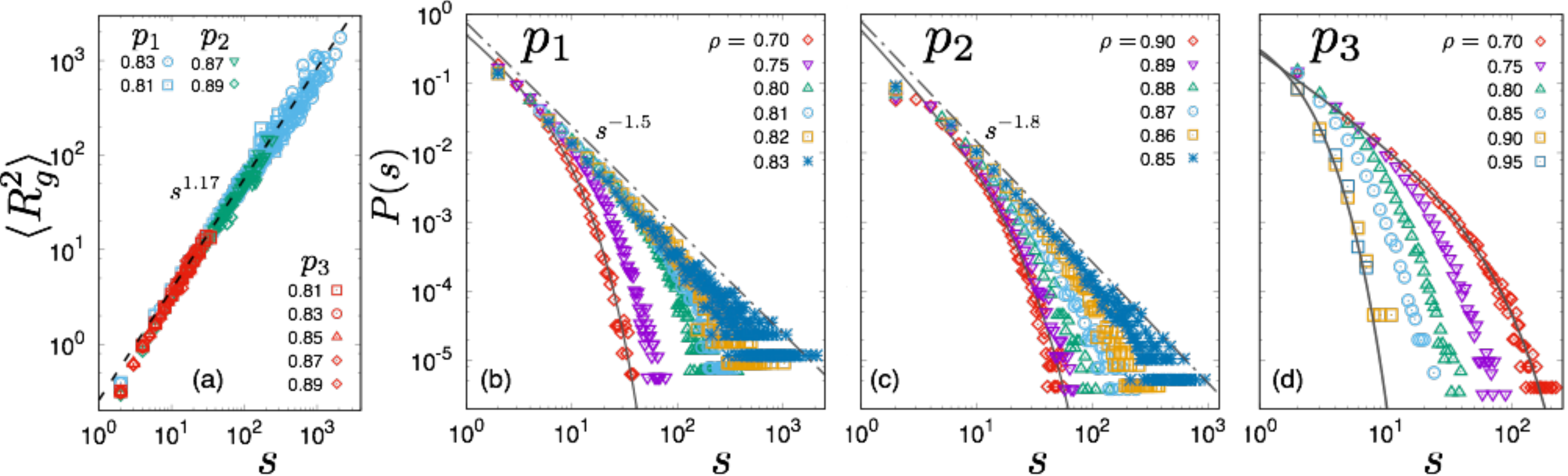}
\end{center}
\caption{
Distribution $P(s)$ of clusters of mass $s$ for different classes of particles over a range of fluid densities. (a) The cluster mass distribution for $p_1$ particles is nearly exponential (shown by {\em solid} line) for low $\rho$, but with increasing $\rho$, it gradually approaches power-law, becoming scale-free at the beginning of coexistence regime, $\rho=0.83$. The power-law exponent is shown by a {\em dashed-dot} straight line in log-log scale. (b) $p_2$ clusters show a complementary feature of $p_1$ clusters as a function of $\rho$. (c) The $p_3$ clusters never percolate over the $\rho$ range studied for our model system. We quantify the size of non-percolating clusters by computing radius of gyration $R_g$ of the clusters as a function of their mass, $s$. (d) $\langle R_g^2\rangle$ showing a near linear dependence on $s$ in log-log scale is observed for clusters from each population for all $\rho$. Only small clusters $s<10$ seem to deviate from this asymptotic scaling behavior.
}
\label{clstr}
\end{figure*}

First, the radial distribution $g(r)$ of particles belonging to each population is computed and presented for different $\rho$ (Fig.\ref{grlkl}). Note that the asymptotic value of $g(r)$ is mostly less than unity. This value indicates the number density of the specific population as that has been used to normalize $g(r)$ instead of the bulk $\rho$ of the system. $p_1$ population shows a $g(r)$ typical to crystalline phase at high $\rho$. This is expected as the particles mostly have $6$ neighbors, characteristic of a hexagonal crystal symmetry in 2D. With decreasing $\rho$, the $p_1$ population decreases and corresponding $g(r)$ shows liquid-like behavior which shows having $6$ neighbors does not ensure crystallinity, pointing to the limitation of {\em coordination number picture}. The $p_2$ population shows consistent liquid-like behavior for all $\rho<0.87$. This further strengthens our view of liquid as a distorted $5$-fold structure since $p_2$ represents the population of $5$-coordinated particles. At $\rho=0.87$, a secondary peak in between the first and second one abruptly appears in $g(r)$ for $p_2$. This feature is weakly shared by $g(r)$ of $p_3$ for the same $\rho$. While this feature definitely indicates local particle crowding, we believe it results from special arrangements of particles with similar area, but different coordination.  Particle crowding relaxes as $\rho$ decreases pushing the peak to large $r$. For low $\rho$, the number of particles constituting the $p_3$ population increases resulting in a split double peak broader than the same for high $\rho$ triangular crystals. We tentatively associate this with local square arrangement of particles, remembering the dominance of $4$-coordinated particles in $p_3$. A detailed analysis of pair correlation of $p_3$ particles in much larger system, especially, showing square-triangle structural transition, may provide further insight and we plan such an analysis in the future.

Next, we carry out a cluster analysis for particles in each population. If two particles belonging to the same population also happen to be each others neighbors, they are considered to form a cluster. Clusters of different particle number or mass $s$, are then identified and their size is quantified by their radius of gyration, $R_g$, as $R_g^2=\frac{1}{2s}\sum(r_i-r_j)^2$ where both $i$ and $j$ particles belong to the same cluster. The average radius of an isotropic cluster is known to follow the scaling relation: \cite{degenes} $\langle R_g^2\rangle\sim s^{2\nu}$ for any finite-size non-spanning cluster of mass $s$, where $\nu$ is the mass scaling exponent, which is roughly equal to $0.6$. \cite{redner} Fig.\ref{clstr}(a) showing a plot of $\langle R_g^2\rangle$ as a function of $s$ for all non-spanning clusters from all populations for all $\rho$ demonstrates an excellent agreement with this theoretical prediction. A slight anisotropy is observed for only small clusters $s<10$, mostly belonging to $p_3$, as they show small deviation from the prediction. For any self-organized finite-size structure, the size is known to follow a distribution: \cite{perco} $P(s)\sim s^{-\alpha}\exp(-s/s_0)$ where $s_0$ denotes the characteristic size and $\alpha$ is the {\em Fisher exponent}.\cite{fisher} $P(s)$ changes from exponential to power-law behavior if the clusters undergo percolation, i.e., there exists at least one cluster that spans the system. This is a geometric transition which marks the loss of any characteristic size of spatial organization within the system as it becomes scale free. No such transition is observed for $p_3$ clusters possibly because of relative rare abundance of $p_3$ population for all $\rho$. (Fig.\ref{clstr} (d)) Both $p_1$ and $p_2$ clusters, however, undergo such a transition but in a complementary way to each other as a function $\rho$. Both the $p_1$ and $p_2$ clusters are exponentially distributed in the liquid and the solid phase, respectively, and both of these distributions gradually develop a power-law behavior with increasing (decreasing) $\rho$ and finally, becomes scale-free within the coexistence regime; see Figs.\ref{clstr} (b) and (c). We find that the power-law exponent $\alpha$ computed for both classes of particles are smaller than the value expected from standard geometrical percolation theory; i.e., $\alpha=187/91$ in two dimensions. \cite{percorev} This is not unexpected since correlations in the association of the particles are known to alter this exponent as found, for example, in directed \cite{dperco} and explosive \cite{eperco} percolation processes. The populations are clearly correlated and dynamic, i.e., particles belonging to one population at an instant become part of another population later in time. The dynamic nature of this structural heterogeneity prompts us to think of the large clustering of different structural domains in terms of a thermo-reversible self-assembly process where total mass is conserved instead of a purely geometric case. Nevertheless, current analysis points to a fundamental dependence between structure and thermodynamics of soft materials that demands further careful and rigorous study.

\subsection{Average dynamics of local geometric fluctuations}
\begin{figure}[h!]
\begin{center}
\includegraphics[width=0.95\linewidth]{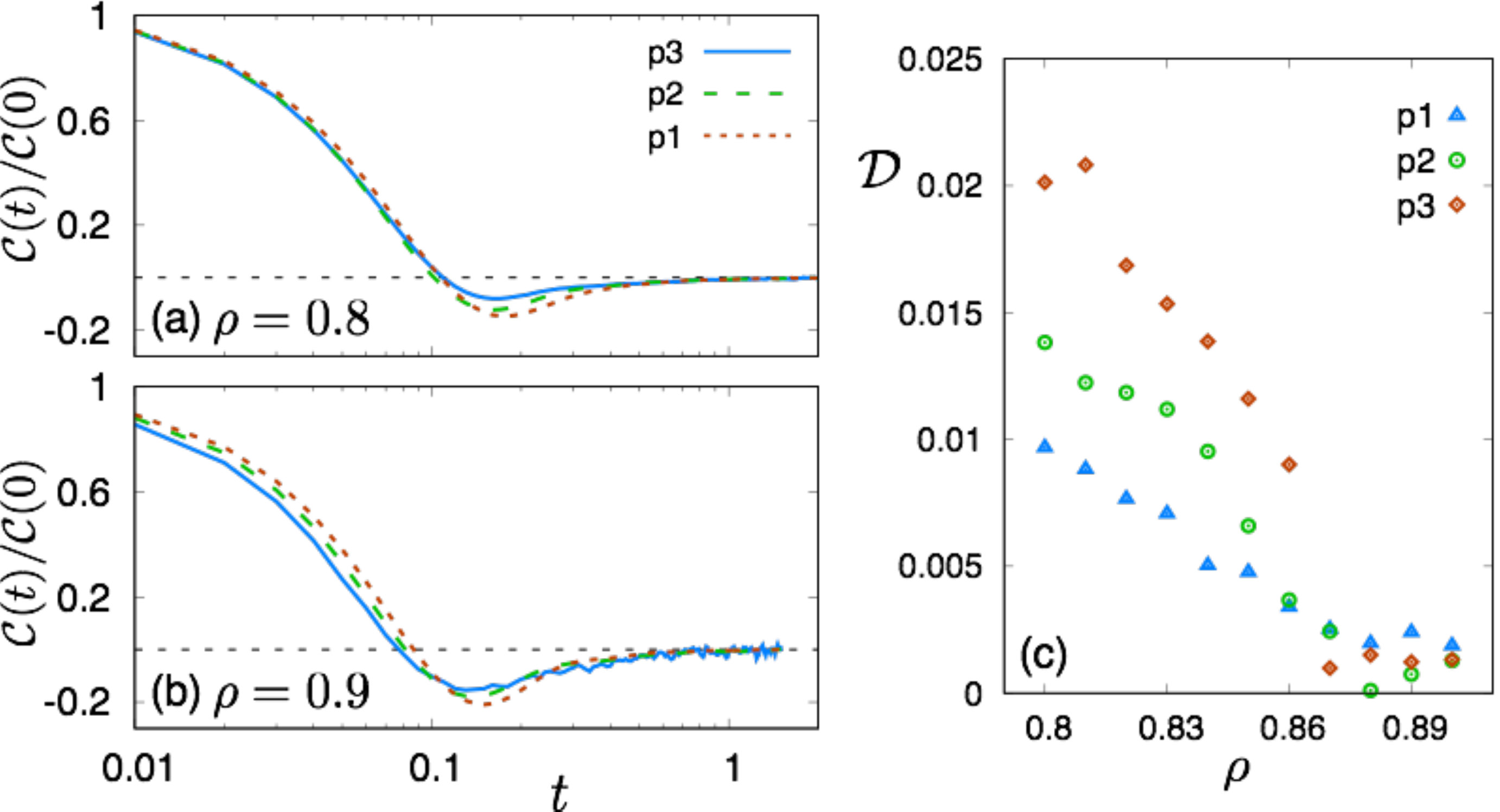}
\end{center}
\caption{
Velocity auto-correlation function $C(t)$ for particles within each population, (a) liquid, $\rho=0.80$ and (b) solid, $\rho=0.90$. $C(t)$ for $p_1$ ({\em solid blue}), $p_2$ ({\em dashed green}) and $p_3$ ({\em dotted red}) appear to be different from each other resulting in a change in diffusivity $\mathcal{D}$. (c) The $\rho$ dependence of $\mathcal{D}$, obtained from integrating $C(t)$, for these populations maintain well-separated $\mathcal{D}$ values within the liquid phase but for large $\rho$, $\mathcal{D}$ becomes similar for each class of particles as the whole system becomes a crystal.
}
\label{vacf}
\end{figure}
Finally, we focus on examining the dynamics of individual population with the aid of the velocity autocorrelation function (VAF) of the constituent particles of each structural population. The VAF, $C(t)=\langle{\bf v}_i(t)\cdot{\bf v}_i(0)\rangle$, provides important information of an interacting system by correlating the velocity ${\bf v}(t)$ of $i$-th particle at time $t$ with its initial velocity ${\bf v}_i(0)$ chosen arbitrarily. On a short time scale, the particles have not had time to collide with the neighbors and their motion is dominated by their inertial motion. As time progresses, the particles start colliding with surrounding particles and thus they `feel' the strong repulsive interactions of these neighboring particles which forces the particle to change its course, resulting in a negative correlation. Due to dynamical rearrangement of all particles, the VAF asymptotically reaches zero marking the onset of diffusive behavior. Over this time scale and beyond, particles are no longer correlated with their initial velocity. Two representative cases, presented in Fig.\ref{vacf} (a) $\rho=0.8$ ({\em liquid}) and (b) $\rho=0.9$ ({\em solid}), show the diffusion time scale is of the order $\tau$ (unit time) for each of the structural populations. Over shorter timescales ($t<\tau$), an observable difference is evident among the normalized VAF for different populations in both cases. The distinction seems to be more prominent in crystal than in the liquid phase. Employing the Green-Kubo relationship, $\mathcal{D}=\int_0^\infty C(t) dt$, we compute the diffusivity $\mathcal{D}$ for each population separately. When plotted against $\rho$, (Fig.\ref{vacf}(c)) $\mathcal{D}$ for $p_1, p_2$ and $p_3$ turns out to be well separated in the liquids ($\rho<0.83$). The difference among them drastically drops down for the systems showing coexistence and crystalline phase as $\mathcal{D}$ itself becomes very small.  This result can be rationalized by the following physical picture: in dense systems, each particle's motion is highly restricted by the closely packed local environment. The positive and negative correlation due to its forward and backward motion become roughly same resulting into a small integral value. In contrast, particles within less dense systems enjoy more free space to move showing larger positive VAF over its negative counterpart occurring from occasional obstruction by their neighbors. This is consistent with our traditional intuition about the caging dynamics in liquids and solids. Our results further confirm the existence of structural dynamic heterogeneity in equilibrium systems by identifying three different structural populations with three distinct average dynamic properties. This evidence of a direct structure-dynamics relationship is promising for developing a theory aimed at understanding the ultimate cause of dynamic heterogeneity. We note that larger system size is required for better resolution data in order to understand the structural population-wise diffusivity in crystals, since both $p_2$ and $p_3$ only occur in tracer amounts within the crystalline phase. In a separate paper, \cite{Lettppr} we further examine the interrelation of dynamical and structural heterogeneity in terms of mobility and lifetime of different structural populations.

\section{Concluding remarks}
In summary, we present a new way to quantify the local structure of a model of many particles interacting at high density. Specifically, we identify the nearest-neighbors of a particle by a solid-angle based criterion and a non-trivial mapping is established between the number of such neighbors and the area of the immediate neighborhood enclosed by those neighboring particles. This tessellation method is simple, fast and parameter-free as the geometric rules employed only use the positional information of angular arrangement of particles. In principle, the algorithm itself should be applicable to a wide variety of systems with arbitrary polydispersity and composition and even for {\em in situ} characterization of particulate systems such as colloids, nanoparticles, granular media etc. An important aspect of this analysis is that it indicates a direct relationship between structural and dynamic heterogeneity in the fast dynamics regime where both types of heterogeneity persist. This promising approach for characterizing structural heterogeneity and its dynamical consequences is potentially very general, but the final assessment of this method requires the study of many other fluids.

Geometric quantification of local structure as is done by our analysis has put forward a picture of a dense simple liquid as a network of (distorted) five-coordinated motifs in contrast to the common image of liquids as a random organization of particles. Importance of $5$-fold local structures has recently been pointed out in the context of glass formation in two-dimensions. \cite{odagaki} Upon transition to a crystalline phase, six-fold motifs dominate this network as expected. Interestingly, a trace amount of four-fold structures has also been detected within the solid phase which start to grow at the solid to liquid transition and continues to grow within liquid phase with decreasing density. Through a set of rigorous tests, we have established that all three such structural populations maintain distinct average dynamics different from each other. This finding advocates that a dense particulate system (both liquids and crystals) can be viewed as a dynamic admixture of three structural populations in contrast to the usual two-population picture of liquid-solid transition. Since previous work has established that there are distinct types of dynamic heterogeneity corresponding to excessively `mobile' and `immobile' particles, \cite{mobimob} in addition to `normal' particles, there must be at least three classes of local structural environments in dense fluids if dynamic heterogeneity is to have a corresponding structural origin. Our analysis is then not only  concerned with the definition, size and location of slow regions having a locally preferred packing, but also regions in which the packing is frustrated and mobility is relatively high. Further, the systematic changes in both local structures and their dynamics are observed to coincide with the thermodynamic change, namely, liquid-solid phase transition of our model system, the analysis method presented here might be the right step forward for microscopic understanding of the structural origin of dynamic heterogeneity of complex systems such as glassy materials.

The notion of dynamic heterogeneity, imagined as the spatial organisation of transient dynamics, is commonly invoked to explain the anomalous response of complex non-equilibrium systems. The present study shows that similar microscopic processes occur even within an equilibrium system and provides a well-defined measure of structural dynamic heterogeneity in dense fluids having a demonstrated relation to the fluid dynamics. The facile classification of the particles belonging to different structural/mobility classes also allows for a characterization of correlation arising from the configurations of these particles within the fluid and the lifetime of these dynamic clusters. Our observations on this clustering processes suggest that it might be possible to understand phase transitions of particulate systems in terms of self-assembly processes, developed elsewhere to describe the self-assembly of polymers in solutions. \cite{debbie} Constructing a dynamic field theoretic description by using the average lifetime of population and their transition rates can be instructive in this regard which we plan to pursue in future. Finally, the phenomenological structure-dynamics-thermodynamics interrelation revealed through our study encourages to accept and further explore the importance of local geometry over the traditional mean-field approaches based on local density, coordination number, etc. or microscopic approaches based on topological defects, to understand dynamic heterogeneity and its consequences. While extensive testing this technique on complex systems and for higher spatial dimensions remains to be performed, the insights gained from the present study should be broadly applicable to understand the dynamical response of soft matter systems.

As a final note, we mention that recent computational methods using machine learning have identified fluctuations in local fluid structure that correlate with local mobility fluctuations occurring on a very short timescale \cite{machinelearning, Cubuk} as in the present work. In particular, these machine learning studies of dense fluids identify an abstract quantity referred to as `softness' which was  a posteriori shown to be related to the number of particles about a particle distance defined by the first peak in the pair correlation function $g(r)$, a measure of the local coordination number. Particles having less neighboring particles than average are referred to as being `soft' the field of values of relative softness defined in this way was defined, visualized, quantified, and then correlated with a range of material properties. Our method of characterizing dynamic structural heterogeneity is likewise based on defining a local effective coordination number, although we utilize a direct tessellation process rather than machine learning to construct the field values of the local coordination number. We have also found that the persistence time of particles in particles in coordination number defined structural states varies with the relative magnitude of the local coordination number (`softness'). Future work should investigate whether a quantitative relation exists between our method of estimating structural heterogeneity and the structural heterogeneity measures determined from machine learning. It is well known that the Debye-Waller factor $\langle u^2\rangle$ provides an experimental measure of local material `softness' \cite{Zaccai, dw3} on a timescale on the order of ps, albeit a dynamical measure rather than structural one, so it would also be interesting to determine the extent to which local coordination number correlates with this experimentally measurable softness measure. Recent work has demonstrated a quantitative relationship between and the structural relaxation time $\tau_\alpha$ in fluids ranging from simulated glass-forming polymer melts, \cite{dw3} metallic glass-forming melts \cite{jack.jstatmech.16} and water \cite{Horstmann} so that enhanced understanding of the structural origin of $\langle u^2\rangle$ is promising for the development of a predictive model of the dynamics of glass-forming liquids based on structural information.

\begin{acknowledgments}
TD acknowledges support under the Cooperative Research Agreement between the University of Maryland and the National Institute of Standards and Technology Center for Nanoscale Science and Technology, Award 70NANB10H193, through the University of Maryland. 
\end{acknowledgments}

\end{document}